\documentclass{article}

\usepackage[english]{babel}

\usepackage[letterpaper,top=2cm,bottom=2cm,left=3cm,right=3cm,marginparwidth=1.75cm]{geometry}

\usepackage{amsmath}
\usepackage{graphicx}
\usepackage[colorlinks=true, allcolors=blue]{hyperref}

\title{Enlightening the chemistry of infalling envelopes and accretion disks around Sun-like protostars: \\ 
the ALMA FAUST project}
\author{\bf C. Codella\,$^{1,2,*}$, C. Ceccarelli\,$^{2}$, C. Chandler\,$^{3}$, N. Sakai\,$^{4}$, S. Yamamoto\,$^{5}$,\\  \bf and the FAUST team\,$^{6}$}

\begin{document}
\maketitle

\noindent
$^{1}$ INAF, Osservatorio Astrofisico di Arcetri, Largo E. Fermi 5, I-50125, Firenze, Italy\\
$^{2}$ Univ. Grenoble Alpes, CNRS, IPAG, 38000 Grenoble, France\\
$^{3}$ National Radio Astronomy Observatory, PO Box O, Socorro, NM 87801, USA\\
$^{4}$ RIKEN Cluster for Pioneering Research, 2-1, Hirosawa, Wako-shi, Saitama 351-0198, Japan\\
$^{5}$ Department of Physics, The University of Tokyo, 7-3-1, Hongo, Bunkyo-ku, Tokyo 113-0033, Japan\\
$^{6}$ http://faust-alma.riken.jp

\begin{abstract}
The huge variety of planetary systems discovered in recent decades likely depends on the early history of their formation. In this contribution
we introduce the FAUST Large Program, which focuses specifically on the early history of Solar-like protostars and their chemical diversity at scales
of $\sim$ 50 au, where planets are expected to form. In particular, the goal of the project is to reveal and quantify the variety of chemical
composition of the envelope/disk system at scales of 50 au in a sample of Class 0 and I protostars representative of the chemical
diversity observed at larger scales. For each source, we propose a set of molecules able to:
(1) disentangle the components of the 50-2000 au envelope/disk system; (2) characterise the organic complexity in each of them;
(3) probe their ionization structure; (4) measure their molecular deuteration. The output will be a homogeneous database of
thousands of images from different lines and species, i.e., an unprecedented source-survey of the chemical diversity of Solar-like
protostars. FAUST will provide the community with a legacy dataset that will be a milestone for astrochemistry and
star formation studies.
\end{abstract}

{\it Frontiers Astronomy and Space Sciences, in press}  


\section{Introduction: the roots of FAUST}

Planets are a common product of the star formation process and there is an incredible variety of planetary
systems in the Galaxy (e.g., http://exoplanet.eu/), very different from the Solar System. The origin of such
diversity, both in physics and chemistry, probably resides in the earliest history of the system formation,
namely what happens during the protostellar phases. 
The low-mass star formation is the complex process transforming a diffuse atomic cloud into
first a dense molecular cloud and eventually into a Sun-like star surrounded by its planetary system. 
Meanwhile, the chemical composition of the gas involved in this process increases its complexity,
from mostly atomic clouds to the so-called interstellar Complex Organic Molecules (iCOMs), i.e. 
species with at least 6 atoms (e.g. methanol, CH$_3$OH), which can be considered as a brick
to build a pre-biotic chemistry (e.g. Ceccarelli et al. 2017, Herbst et al. 2009, Caselli \& Ceccarelli 2012, J{\o}rgensen et al. 2020, and references therein).
Nowadays, evidence is mounting that the first steps of the process, 
namely when the protostar is in the so-called Class 0 ($\sim$ 10$^4$ yr) and I ($\sim$ 10$^5$ yr) source phases
(e.g. Andr\'e et al. 2014, and references therein), 
are crucial for the future of the nascent planetary system. More specifically, a breakthrough result 
has been provided by the ALMA 
(Atacama Large Millimeter/submillimeter Array) interferometer, which provided images of rings and gaps 
(e.g. Sheehan et al. 2017, Fedele et al. 2018, Andrews et al. 2018) 
in the dust distribution around objects with an age less than 1 Myr. 
This supports that planet formation starts earlier than the classical protoplanetary
stage (Class II; $\sim$ 10$^6$ yr).
These findings then support the importance of investigating the chemical complexity
associated with protostars younger than 1 Myr.

In a nutshell, a Sun-like protostar is accreting its mass through a disk rotating 
along the equatorial plane. The disk, expected with a radius less than 100 au, is perpendicular to fast ($\sim$ 100 km s$^{-1}$) 
jets removing the angular momentum excess 
(see e.g. Frank et al. 2014, Lee et al. 2017b). 
In turn, the disk 
is fed by a large scale ($\geq$ 1000 au) rotating and infalling molecular envelope. 
The chemical composition of the envelopes
surrounding Solar-like Class 0 and I protostars on large scales (100 -- 2000 au) can be very different. 
Two distinct classes have been discovered: (1)  the hot corinos (Ceccarelli et al. 2007)  and the WCCC (Warm Carbon Chain
Chemistry) sources (Sakai et al. 2013). Hot corinos (as e.g. IRAS16293-2422)  are compact ($\geq$ 100 au), and dense ($\geq$ 10$^7$ cm$^{-3}$) regions, 
where the temperature is warm enough ($\geq$ 100 K) to thermally evaporate the frozen dust mantles.
As a consequence the gas phase is chemically enriched in iCOMs, due to either direct release from 
dust mantles or formed in gas phase using simpler molecules from the mantles. 
On the other hand,  WCCC sources (as e.g. L1527) are deprived of such iCOMs
and enriched with unsaturated carbon-chain species, such as HC$_{\rm 2n+1}$N, C$_{\rm n}$H, and C$_{\rm n}$H$_2$
(Higuchi et al. 2018, and references therein).

In this context, breakthrough open questions are (i) whether such diversity
is also present in the inner envelope/disk system ($\sim$ 50 au), and (ii) what molecules are passed from the
large-scale envelope ($\sim$ 2000 au) to the disk in which planets, comets and asteroids form. Yet, this is crucial
to know because very likely the chemical composition and fate of the future planetary systems depend on the
chemical class to which the original protostar belonged. The aim of this paper is to introduce 
the ALMA (Atacama Large Millimeter/submillimeter Array) 
Large Program (LP) FAUST (Fifty AU STudy of the chemistry in 
the disk/envelope system of Solar-like protostars; 
http://faust-alma.riken.jp), focused on astrochemistry
of Class 0/I sources imaged at the Solar System spatial scale. 

\section{The FAUST project}

\subsection{Context and main goals}

The goal of FAUST is to characterize the gas chemical composition of  
the protostellar system, associated with several physical components at work to build a Sun-like
star from a rotating and infalling envelope.
FAUST builds up on previous surveys of the chemical composition of the Sun-like youngest protostars at larger spatial scales, such 
as ASAI (https://www.oan.es/asai/, Lefloch et al. 2018), and SOLIS (https://solis.osug.fr/, Ceccarelli et al. 2017), which covered the large envelopes at $\geq$ 5000 au and the transition between the large-scale envelope and the circumstellar disk at $\leq$ 5000 au and $\geq$ 100 au, respectively.
The FAUST approach is to identify three zones (see Fig. 1)
as learnt from previous ALMA studies
(e.g. Sakai et al. 2014b, 2017): 

\begin{figure}
\centering
\includegraphics[angle=90,width=1.0\textwidth]{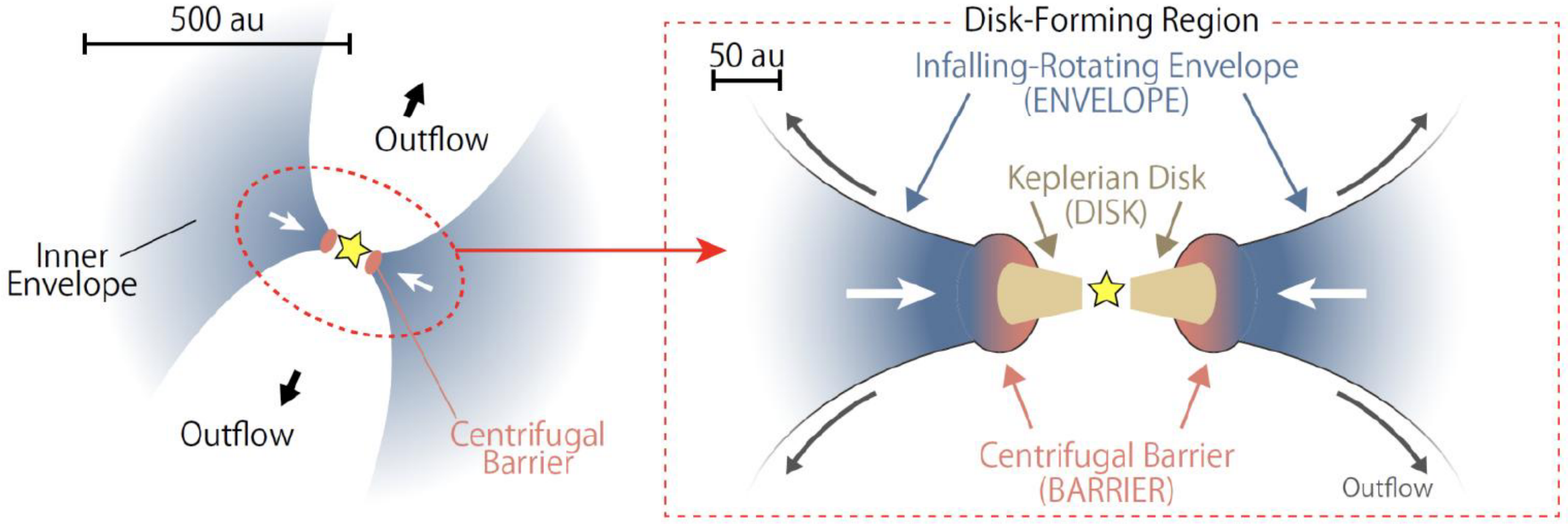}
\caption{Schematic structure of the region around a Solar-like Class 0 and I protostar,
adapted from Oya et al. (2016).
Three main zones can be identified (see Sect. 2): the infalling and rotating envelope,
the centrifugal barrier, and
the rotating and accreting disk.}
\end{figure}

\begin{itemize}
\item
ENVELOPE: Here we mean the infalling-rotating 
envelope on scales of a few 100 au. The gas chemical composition evolves from that at large ($\sim$ 2000 au) 
scales because of the heating from the central object, which sublimates the grain mantles (Ceccarelli et al. 1996). 
Besides, the gas close to the outflow cavity wall might also be exposed to the UV- and X-ray photons 
from the central object or be affected by mild transverse shocks (Stauber et al. 2005),
with a consequent chemical enrichment of the gas.

\item
BARRIER: The gas transits a
centrifugal barrier, on scales of about 50 au, before entering the disk. The gas chemical composition 
may be drastically affected by the low velocity ($\sim$ 1 km s$^{-1}$) 
shock at the centrifugal barrier, as grain mantles may be at least 
partially 
liberated into the gas phase and the gas heated and compressed (Sakai et al. 2014ab, Oya et al. 2016). 
As a matter of fact, the IRAS 04386+2557 protostar, in the L1527 Taurus core, 
can be considered the archetypical
edge-on disk where (Sakai et al. 2014a, 2017) revealed, for the first time, 
an increase of SO abundance at the radius of the centrifugal barrier.
Another enlightening case is represented by the HH212 pristine jet/disk system, located
in Orion B. The disk, edge-on also in this case, shows iCOMs-rich rotating rings,
possibly associated with the centrifugal barrier 
with a radius of about 40 au (Lee et al. 2017c, 2019).  

\item
DISK: At scales smaller than about 50 au, 
the gas settles in the rotationally-supported disk, where the gas chemical composition is expected to be 
stratified (e.g. Aikawa et al. 1999, Walsh et al. 2015) and affected by the 
dynamics and the dust coagulation (e.g. Zhao et al. 2016, Ilee et al. 2017). 
Indeed, very recently, Podio et al. (2020) observed with ALMA 
the evolved Class I source
IRAS 04302+2247, where the bulk of the envelope has been dispersed: 
the molecular (CO, CS, H$_2$CO) emission
is vertically stratified on a scale of 50-60 au 
(see also van’t Hoff et al. 2020).
A breakthrough result would be to reveal a chemical
stratification in early protostellar disks still fully embedded in the envelope.
Furthermore, the observations of protostellar disks are also challenged by
the presence of jets
and disk winds, which are powered at similar scales (e.g. Lee et al. 2017b, Tabone et al. 2017). 
For a proper study of disks it is mandatory to reveal also  
the fast jet flowing perpendicular to the disk equatorial
plane using e.g. a standard tracer as SiO (e.g. Codella et al. 2019). 
Finally, at these small scales dust opacity increases, which strongly affects the emission
spectrum from the gas (see below). 
\end{itemize}

Therefore, the three zones are expected to possess distinct physical and chemical properties, 
likely varying from source to source and/or depending on the star forming regions. 
The goal of FAUST is to disentangle the three zones with the help 
of their kinematic signature and ALMA line images at 50 au spatial resolution. Our pioneering studies 
showing the proof-of-concept have already been conducted for a few sources and a few molecules 
(e.g. Ceccarelli et al. 2010, Oya et al. 2016, Codella et al. 2017):
time is now ripe for a systematic study of much more sources and much more species, 
via an ALMA LP. In summary, the aim of FAUST is to reveal and quantify the variety of the 
chemical composition of the envelope/disk system of Solar-like Class 0 and I protostars. Such chemical 
variety will add a new dimension to the diversity of planetary systems, and we expect that it will 
have a substantial impact on studies of planet formation and the origin of the Solar System.

\subsection{Sources and molecular lines}

From a kinematical point of view, the envelope/disk system can be divided into three different zones.
Although they could, in principle, be studied kinematically by observing CO
and its isotopologues lines, but as a matter of fact only rarer species 
provide a much more powerful diagnostic tool, because they are differentially enhanced in the three zones
thanks to chemical composition changes 
(e.g. Sakai et al. 2014a, 2017, Oya et al. 2016).  
FAUST will simultaneously use kinematics and
chemistry to fully resolve the complexity of the envelope/disk system.
Four groups of species have been selected to probe different topics: 

\begin{enumerate}
\item
ZONE PROBES: c-C$_3$H$_2$, CS, CH$_3$OH, SO, SiO, H$_2$CO, C$^{18}$O, and HC$_3$N.
We select the following species to disentangle each zone:
Envelope: c-C$_3$H$_2$, CS. These species are present in the infalling-rotating envelope, with different
abundances for different sources, but their abundance significantly drops in the other two zones, so that
they are specific in probing the ENVELOPE zone in different sources.
Barrier: CH$_3$OH, SO, SiO. Their abundance is enhanced in the weak shocks at the centrifugal barrier.
%Depending on the source chemical type, either of CH$_3$OH or SO is more abundant. 
Narrow ($\sim$ 1 km s$^{-1}$) 
SiO lines 
could trace the release of Si from dust mantles
(Guillet et al. 2011, Lesaffre et al. 2013) at the centrifugal barrier, 
whereas broad ($\geq$ 10 km s$^{-1}$) SiO emission is expected
to trace fast jets (associated with high-velocity shocks)  perpendicular to the disk. 
Disk: H$_2$CO, C$^{18}$O, and HC$_3$N. Given their relatively large abundance in the warm layers of the disks,
lines from these species can probe the inner disk, via their high-velocity components.

\item
MOLECULAR COMPLEXITY PROBES: CH$_3$OH, NH$_2$CHO, CH$_3$CHO, CH$_3$OCH$_3$, and HCOOCH$_3$.
A major goal of FAUST is measuring the organic complexity in the disk/envelope system, as it might be
inherited at later stages by the nascent planetary system (Zhao et al. 2016). Based on previous studies, five species are
particularly important to identify the organic diversity: methanol (CH$_3$OH), formamide (NH$_2$CHO),
acetaldehyde (CH$_3$CHO), dimethyl ether (CH$_3$OCH$_3$) and methyl formate (HCOOCH$_3$). They are
predicted to have a different chemical origin, with methanol a grain-surface product, formamide likely a
gas-phase product (Codella et al. 2017, Skouteris et al. 2017), 
while for acetaldehyde, dimethyl ether and methyl formate the chemical
synthesis route is largely debated
(e.g. Garrod et al. 2008, Balucani et al. 2015).

\item
GAS IONIZATION PROBES: H$^{13}$CO$^+$, DCO$^+$, and N$_2$H$^+$.
The degree of ionization in the inner 100 au envelope is a very important parameter for any theory of
planet formation (e.g. Balbus \& Hawley 1998). Yet, this quantity is very poorly known at these scales. 
Several processes can
affect it, from an inner source 
of energetic particles and/or X-rays (Stauber et al. 2005) to the growth of dust grains (Zhao et al. 2016).
Here we propose to use molecular ions, following a previously used methodology, to estimate the gas
ionization, such as the H$^{13}$CO$^+$/DCO$^+$ 
and H$^{13}$CO$^+$/N$_2$H$^+$ abundance ratios.
The first ratio
measures the ionization in cold (smaller than $\sim$ 30 K) 
and dense ($\geq$ 10$^4$ cm$^{-3}$) gas (e.g. Caselli et al. 2008), 
while the second one in warm ($\geq$ 40 K) 
and, again, quite dense (up to $\sim$ 10$^7$ cm$^{-3}$) gas (e.g. Ceccarelli et al. 2014a).

\item
DEUTERIUM-BEARING SPECIES: c-C$_3$HD, N$_2$D$^+$, HDCO, D$_2$CO and CH$_2$DOH
D-bearing species are a powerful diagnostic tool to study the physical 
conditions at present and in the past (Ceccarelli et al. 2014b). Specifically, species like c-C$_3$H$_2$ (one of the infalling-rotating envelope probe) 
and N$_2$D$^+$ are present-day
products, whereas H$_2$CO and CH$_3$OH, being major components of the grain mantles, were mostly formed
during the prestellar phase and introduced in the present-day gas by the presence of the accreting inner
object. The respective deuterated counterparts provide, therefore, a precious tool to understand the
physical conditions when these species are/were deuterated. In addition, molecular deuteration provides a
sort of Ariadne's thread that links the ISM to the Solar System history
(Ceccarelli et al. 2014b). 
\end{enumerate}

We selected a representative sample of sources
that are known to exhibit a wide chemical composition diversity, based on the available large-scale
observations. The following two criteria have been adopted:
(1) Chemical diversity: Since, so far, the major chemical diversity is represented by WCCC and hot corino
sources, we select sources that represent a continuous variation of the abundance ratio of species
characteristic of these two classes. Specifically: (i) CH$_3$OH 
can be regarded as a proxy of the hot corino
species, because it is a crucial organic molecule, considered a parent species of larger iCOMs; (ii) small
hydrocarbons such as C$_2$H or c-C$_3$H$_2$ 
are characteristics of WCCC sources. The selected sources are therefore associated
with measured abundance ratios of these species varying by two orders  
of magnitude, covering 
the two extremes of hot corino and WCCC sources
(Higuchi et al. 2018). 
(2) Distance and previously studied sources: We select nearby 
Class 0 and I sources with distance $\leq$ 250 pc and a 
bolometric luminosity $\leq$ 25 $L_{\rm \odot}$. 

\begin{table*}
\caption{List of the frequency windows and of the species covered by the 
selected backends, divided in four groups 
(see Sect. 2.2).}
\begin{tabular}{cccccc}
\hline
\multicolumn{1}{c}{Setup} &
\multicolumn{1}{c}{Frequency} &
\multicolumn{1}{c}{Zones} &
\multicolumn{1}{c}{Molecular} &
\multicolumn{1}{c}{Gas} &
\multicolumn{1}{c}{Molecular} \\ 
\multicolumn{1}{c}{ } &
\multicolumn{1}{c}{(GHz)} &
\multicolumn{1}{c}{ } &
\multicolumn{1}{c}{complexity} &
\multicolumn{1}{c}{ionisation}  &
\multicolumn{1}{c}{deuteration} \\ 
\hline
1 & 214.0-219.0 & c-C$_3$H$_2$, CH$_3$OH, SO & CH$_3$OH, NH$_2$CHO & DCO$^+$ & N$_2$D$^+$, D$_2$CO \\
  & 229.0-234.0 & SiO, H$_2$CO, C$^{18}$O      & CH$_3$CHO, HCOOCH$_3$ &       &             \\
\hline
2 & 242.5-247.5 & c-C$_3$H$_2$, CS  & CH$_3$OH, NH$_2$CHO & H$^{13}$CO$^+$ & HDCO, CH$_2$DOH \\
  & 257.5-262.5 & CH$_3$OH, SO      & CH$_3$CHO, HCOOCH$_3$ &       &             \\
  &             &                   &    CH$_3$OCH$_3$      &       &            \\
\hline
3 & 85.0-89.0 & CH$_3$OH, HC$_3$N & CH$_3$OH, $^{13}$CH$_3$OH & N$_2$H$^+$ & c-C$_3$HD \\
  & 97.0-101.0 &                     &                   &       &             \\
\hline
\end{tabular}
\end{table*}

The final list contains 13 sources located in different star forming regions, namely:
L1527, and L1551 IRS5 (Taurus region), IRAS 15398-3359 (Lupus), L483 (Aquila), Elias 29, VLA1623A, 
IRS63, and GSS30 (Ophiucus), NGC1333 IRAS4A, and IRAS4C (Perseus), BHB07-11 (Pipe), RCrA IRS17A (Cr A),
and CB68 (Isolated source). 
We will observe the three frequency settings in Tab. 1. They target lines
from the four groups of molecules described above. 
We require a uniform linear resolution of 50 au to identify the centrifugal barrier, and to 
disentangle the disk from the envelope, namely an angular
resolution of 250-350 mas, depending on the distance of the source, and almost uniform brightness sensitivity.
The high-angular scale of 50 au will be reached using the 12-m antenna ALMA array 
in different configurations. 
Furthermore, given the goal to sample also the molecular envelope, 
Atacama Compact Array (ACA/Morita Array) observations of 7m antennas will be also performed. 
By combining all the visibility data obtained using both the 12m and the 7m antennas array,
FAUST will be able to follow the
physical and chemical changes from the large-scale envelope ($\sim$ 2000 au) to the inner disk/jet system.
For each frequency setting, we allocate
one continuum setting to measure the dust Spectral Energy Distribution (SED) from 3 mm to 1 mm. 
The frequency resolution is set to 0.2
km s$^{-1}$, i.e. the sound speed at 10 K.

\subsection{Data exploitation}

We will use a two-step modeling procedure to extract the information from the data, as follows:
(1) Kinematics and radiative transfer models: The three zones are disentangled by comparing the velocity
structure observed in the various molecular lines with kinematic models 
(Oya et al. 2016). The gas temperature and
density structure, as well as the species column density, will be extracted 
from a multiline non-LTE (when
possible) analysis using 1 and 3D codes (Bisbas et al. 2015, Zhao et al. 2016), taking into account the dust optical depth.
The analysis of the continuum will provide the distribution of the H$_2$ column density.
(2) Astrochemical models: The derived molecular abundances will be compared with our astrochemical
models (Ceccarelli et al. 2018, Wirstr\"om et al. 2018), where the physical structure will be incorporated. From
this analysis we will extract valuable constraints on the 
reactions involved in the formation and destruction
of the observed iCOMs. FAUST members with specific expertise will
contribute to the project with laboratory experiments and theoretical quantum chemistry calculations (Watanabe et al. 2010, Dulieu et al. 2013, Skouteris et al. 2017).
In summary, we will determine (i) the extent to which the large-scale envelope diversity is
conserved in the disk/envelope system and (ii) what molecules are passed from the large-scale envelope to
the disk. In this way, the goal of the FAUST project will, therefore, be achieved.

\section{First FAUST results and perspectives}

The analysis of the FAUST data is in progress. However, some instructive results have been already obtained.
The data were reduced by using a modified version of the ALMA calibration pipeline and an
additional in-house calibration routine (G. Moellenbrock et al.
2021, in preparation) to correct for the system temperature and spectral line data normalization.
Figure 2 reports an example of the zoo of FAUST spectra observed in a portion of the frequency Setup 2 (see Table 1) towards the protostellar coordinates. 
The spectra have been obtained
merging ACA data with those at high-spatial resolution.
The line identification and the discussion of the chemical content is out of the focus of the present paper;
the goal of the spectra sketched in Fig. 2 is to enlight a chemical differentiation,  
indicating the source sample is not biased by some selection criterium. More precisely, four sources (L1551, L483, CB68, BHB07-11 and CB68) are characterised by iCOMs emission, revealing hot corino activity. 
The next step is a careful investigation of weaker lines as well as the inspection of the images obtained
using each array configuration, in order to provide a complete characterisation of the molecular complexity in each source. 

\begin{figure}
\centering
\includegraphics[width=1.0\textwidth]{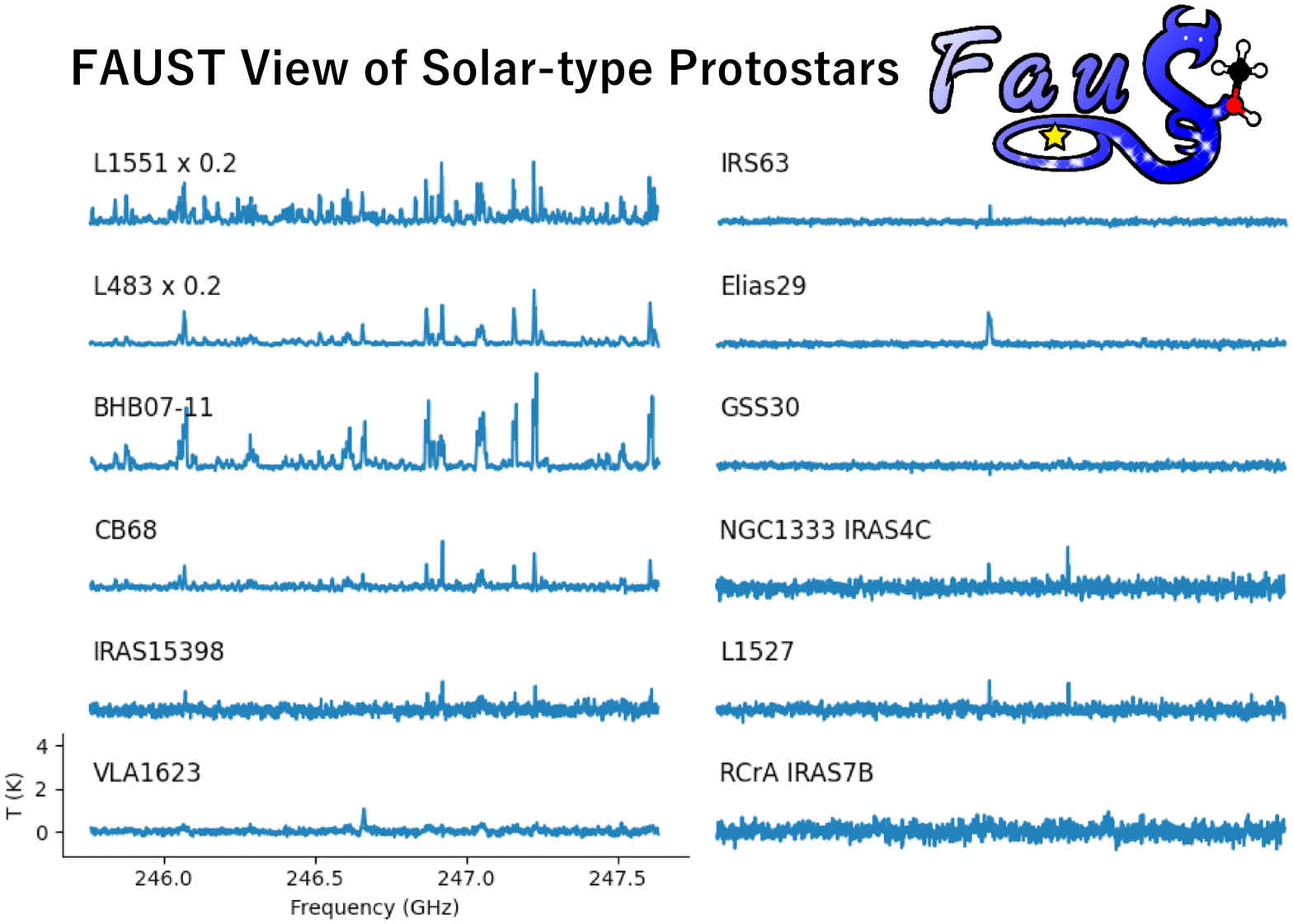}
\caption{Zoo of preliminary spectra observed towards FAUST sources 
(see black labels) in a portion of 2 GHz (245.7--247.7 GHz) of the frequency Setup2 (see Table 1). The FAUST logo
is also reported. The intensity scale is in brightness temperature (K).
The spectra have been obtained merging ACA data with those at 
high-spatial resolution. The identification of the lines is out of the focus of
the present paper. The goal of the sketch is to enlight the observed chemical differentiation, with some sources
characterised by a rich spectra associated with iCOMs emission.}
\end{figure}

An example of what FAUST is able to provide when studying the inner 50 au around protostars is
reported by Bianchi et al. (2020), who imaged the Class I L1551-IRS5 binary system in several species,
namely  methanol (including $^{13}$CH$_3$OH and CH$_2$DOH), 
methyl formate (HCOOCH$_3$), and ethanol (CH$_3$CH$_2$OH).
Figure 3 (Left) shows the dust continuum 1.3 mm emission revealing the binary components, namely the N (northern) and 
S (southern) objects. The two sources are surrounded by a circumbinary disk, 
clearly traced by Cruz-S\'aenz de Miera et al. (2019).  
The 50 au resolution reveals an iCOMs rich hot corino toward the N component (see the HCOOCH$_3$ map in Fig. 3 - Right), 
and, in addition, a possible second hot corino around the S object. The number of 
hot corinos imaged around Class I sources is, so far, limited to a handful number
(e.g. De Simone et al. 2020, Yang et al. 2021).
The discovery of iCOMs emission in L1551-IRS5 is thus instructive in the effort of tracking how the chemical richness 
around Class 0 objects is inherited by later Class I protostars: as a matter of fact, it looks that chemistry does not dramatically change.
Projects on this topic based on statistically reliable samples will be hopefully performed soon.
Figure 3 (Right) shows the map of the velocity peak of the HCOOCH$_3$(18$_{\rm 4,14}$--17$_{\rm 4,13}$A) 
line profile. A clear velocity gradient is revealed
perpendicular to the axes of the protostellar jets (see the arrows in the Left panel) driven by N and S. 
In conclusion, the chemical enrichment occurs in rotating hot corinos around the protostars and/or in
the rotating circumbinary disk.

\begin{figure}
\centering
\includegraphics[angle=90,width=1.0\textwidth]{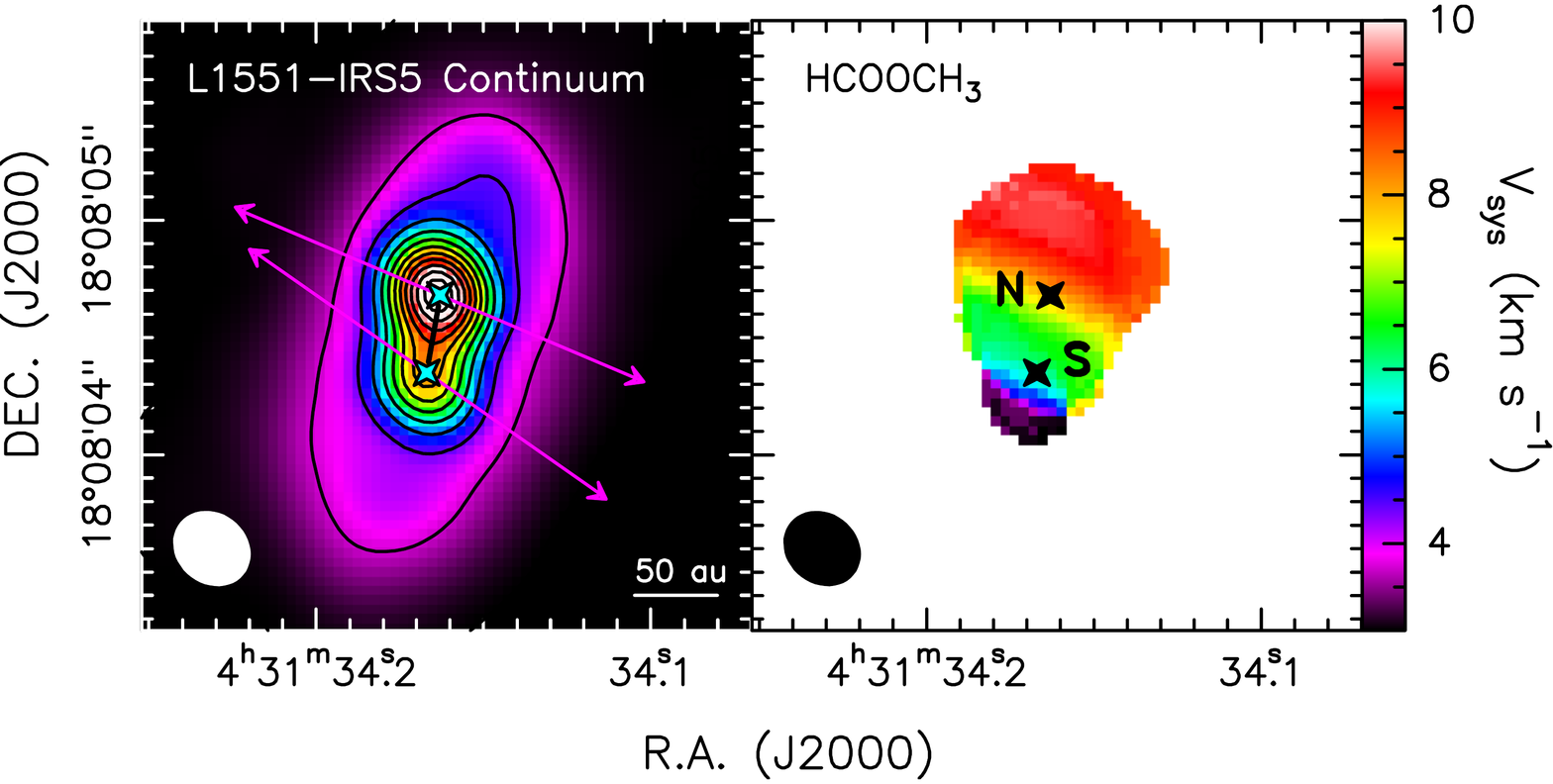}
\caption{The L1551 IRS5 binary source, in Taurus.
{\it Left panel:} 1.3 mm dust continuum emission in colour scale and black contours )Bianchi et al. 2020): the N and S protostars are revealed.
{\it Right panel:} Gas enriched in iCOMs around the IRS5 binary system, imaged on the 50 au scale. 
More specifically,
the intensity-weighted velocity (also called moment 1) map of the HCOOCH$_3$(18$_{\rm 4,14}$--17$_{\rm 4,13}$)A is shown, 
revealing a velocity gradient 
perpendicular to the axes of the protostellar jets (see the arrows in the {\it Left} panel) driven by N and S. 
The systemic velocities are +4.5 km s$^{-1}$ (S object), and +7.5 km s$^{-1}$ (N). 
The chemical enrichment occurs in rotating hot corinos around the protostars and/or in
the rotating circumbinary disk.}
\end{figure}

\begin{figure}
\centering
\includegraphics[width=1.0\textwidth]{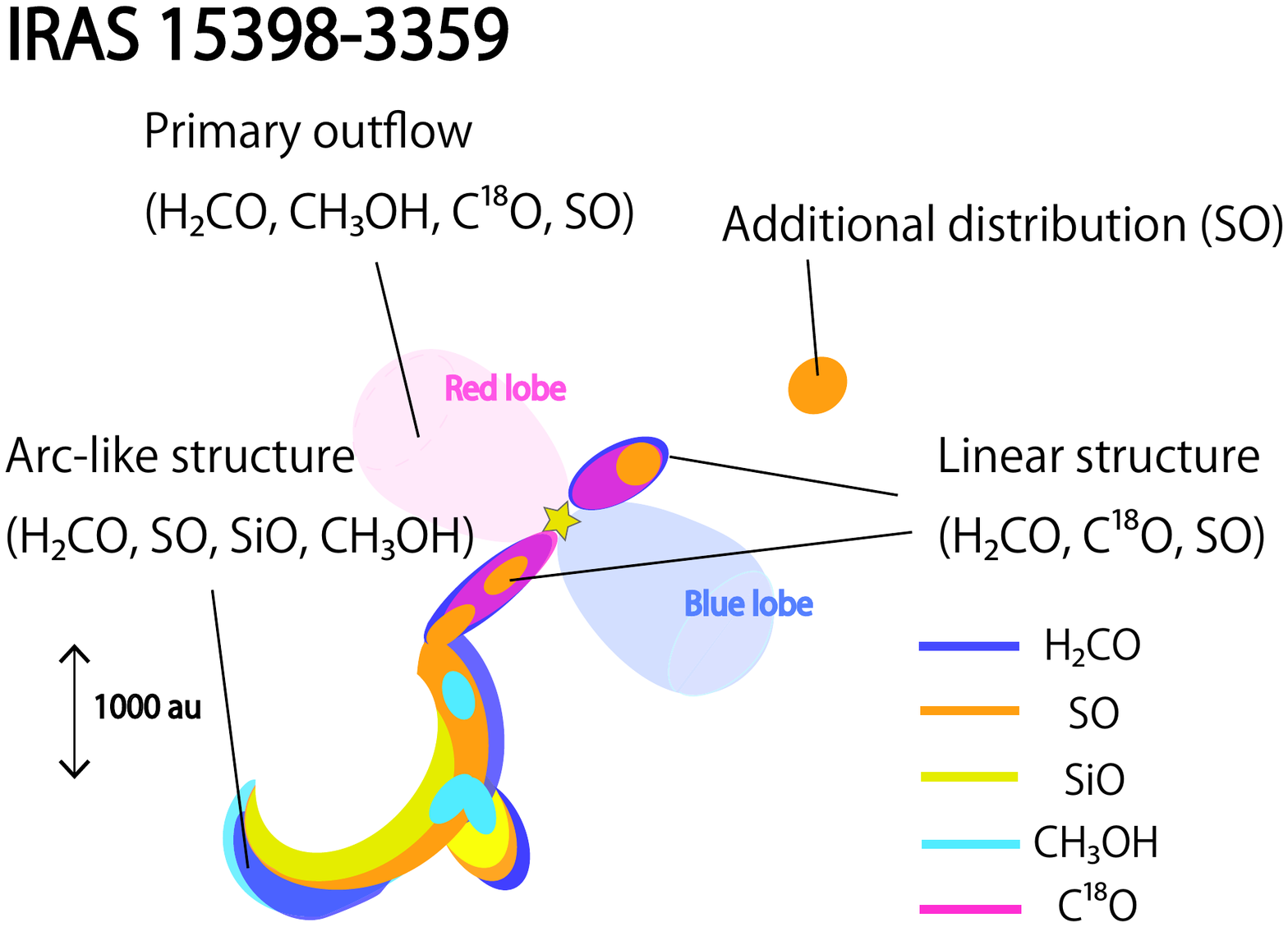}
\caption{Adapted from Okada et al. (2021): 
Schematic picture of the molecular distributions
around the IRAS 15398-3359 protostar located in the Lupus star forming region. 
FAUST observations of emission lines due to
different molecular species disentangle different components on large scale
associated with multiple outflows.}
\end{figure}

Moving to larger spatial scales (from 50 au to 1800 au), FAUST recently published a review of 
the molecular emission in the Class 0 IRAS15398--3359 star forming region in Lupus 
(Okoda et al. 2021).
The FAUST multi-species approach (CO, C$^{18}$O, SO, H$_2$CO, CH$_3$OH, and SiO) allows Okoda et al. (2021) to 
well trace the dense and/or shocked material around the protostellar object.
Beside the already known main outflow located along the NE-SW direction, narrow ($\sim$ 1 km s$^{-1}$) line emission 
of shock tracers such as SiO and CH$_3$OH 
reveals (i) a secondary outflow with an axis normal to the main flow, plus (ii) an arc-like structure related with the secondary outflow
(see the schematic picture in Fig. 4). 
Which is the origin of the secondary outflow? Are the outflows driven by different protostars? 
As a matter of fact, there is no evidence (neither in the FAUST continuum maps)  of a companion of the Class 0 IRAS15398--3359 protostar.
An alternative solution, proposed by Okoda et al. (2021), is that the secondary outflow is a relic of a past reorientation 
of the outflow launched from a single protostar and that the arc-like structure has dissipated the turbulence of the old shocked material. 
In principle, a dramatic change in the direction of the outflow axis could be possible, if the molecular
core hosting the protostar is associated with a non-uniform internal angular momentum, transferred onto the central region via episodic accretion. 

\begin{figure}
\centering
\includegraphics[width=1.0\textwidth]{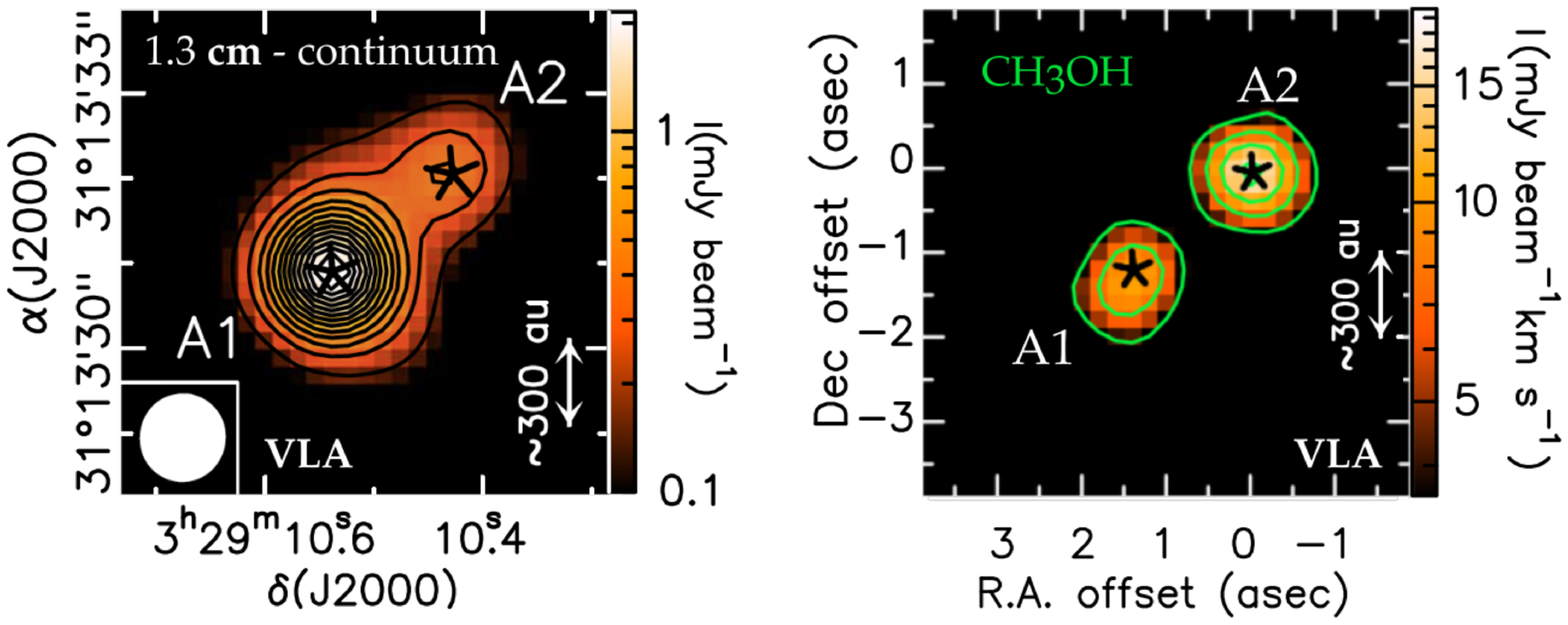}
\caption{{\it Left panel:} NGC1333 IRAS 4A continuum emission map at 25 GHz as 
imaged using VLA; Adapted from De Simone et al. (2020). Two protostars, A1 and A2 are observed. 
{\it Right panel:}
CH$_3$OH(6$_{\rm 2,4}$-6$_{\rm 1,5}$) velocity-integrated maps  
in color scale, revealing for the first time the hot-corino associated with A1, missed
by sub-millimeter surveys due to dust opacity effects.}
\end{figure}

Finally, the FAUST synergy is at work also to prepare complementary observations to those collected with
the ALMA Large Program. More specifically, we started pilot projects in the cm-spectral window using the Very Large Array (VLA) interferometer.
Given the dust opacity at cm-wavelengths is negligible
(e.g. Testi et al. 2021, and references therein), 
the main goal is to evaluate the effects of dust emission on Solar System scales,
associated with both high volume and column densities 
(e.g. Miotello et al. 2014, Galv\'an-Madrid et al. 2018, Galametz et al. 2019) on the 
collected iCOMs images (and derived abundances) at (sub-) mm-wavelengths.
An instructive example is provided by De Simone et al. (2020), who imaged at 1.3 cm at VLA in the FAUST context, the NGC1333-IRAS4A binary system.
One of the two components, 4A1, lacks iCOMs emission when observed at millimeter wavelengths (see Fig. 5), while the other component, 4A2,
is very rich in iCOMs. De Simone et al. (2020) found that, once imaged at 1.3 cm, methanol lines are similarly bright toward 4A1 and 4A2,
proving that both are hot corinos, and not only 4A2. Complementing the ALMA observations with centimeter images can be then considered the future step
which the FAUST synergy will perform. The VLA-ALMA combination will allow us to perform steps ahead in the study of the protostellar environments and 
to understand the ultimate molecular complexity in regions with high column density of material, and consequently high dust opacity.

\section*{Author Contributions}
All authors contributed to the data analysis, discussed the results and commented on the manuscript.

\section*{Funding}
This project has received funding from: 1) the European 
Research Council (ERC) under the European Union's Horizon 2020 research and innovation 
program, for the Project {\it The Dawn of Organic Chemistry} (DOC), 
grant agreement No 741002; 2) the PRIN-INAF 2016 {\it The Cradle of Life - GENESIS-SKA} 
(General Conditions in Early Planetary Systems for the rise of life with SKA); 
3) a Grant-in-Aid from Japan Society for the Promotion of Science (KAKENHI: Nos. 18H05222, 19H05069, 19K14753); 
4) ANR of France under contract number ANR-16-CE31-0013; 5) the French National Research 
Agency in the framework of the Investissements d'Avenir program (ANR-15-IDEX-02), 
through the funding of the {\it Origin of Life} 
project of the Univ. Grenoble-Alpes, 6) 
the European Union's Horizon 2020 research and innovation 
programs under projects {\it Astro-Chemistry Origins} (ACO), Grant No 811312.

\section*{Acknowledgments}
This paper is written on behalf of the whole FAUST team (http://faust-alma.riken.jp) and it makes use 
of the following ALMA data: ADS/JAO.ALMA\#2018.1.01205.L. ALMA is a partnership of ESO 
(representing its member states), NSF (USA) and NINS (Japan), together with NRC (Canada), 
MOST and ASIAA (Taiwan), and KASI (Republic of Korea), in cooperation with the Republic of Chile. 
The Joint ALMA Observatory is operated by ESO, AUI/NRAO and NAOJ. The National Radio Astronomy Observatory 
is a facility of the National Science Foundation operated under cooperative agreement by Associated Universities, Inc.
This is a short text to acknowledge the contributions of specific colleagues, institutions, 
or agencies that aided the efforts of the authors.

\section*{References} 

\hspace{4mm}
Aikawa, Y. and Herbst, E. (1999). Molecular evolution in protoplanetary disks. Two-dimensional distributions and column densities of gaseous molecules. A\&A 351, 233–246 \\
\indent
Andr\'e, P., Di Francesco, J., Ward-Thompson, D., Inutsuka, S.-I., Pudritz, R. E., and Pineda, J. E. (2014). From Filamentary Networks to Dense Cores in Molecular Clouds: Toward a New Paradigm for Star Formation. Protostars and Planets VI, 27–51doi:10.2458/azu uapress 9780816531240-ch002 \\
\indent
Andrews, S. M., Huang, J., P\'erez, L. M., Isella, A., Dullemond, C. P., Kurtovic, N. T., et al. (2018). The Disk Substructures at High Angular Resolution Project (DSHARP). I. Motivation, Sample, Calibration, and Overview. ApJ 869, L41. doi:10.3847/2041-8213/aaf741 \\
\indent
Balbus, S. A. and Hawley, J. F. (1998). Instability, turbulence, and enhanced transport in accretion disks. Reviews of Modern Physics 70, 1–53. doi:10.1103/RevModPhys.70.1 \\
\indent
Balucani, N., Ceccarelli, C., and Taquet, V. (2015). Formation of complex organic molecules in cold objects: the role of gas-phase reactions. MNRAS 449, L16–L20. doi:10.1093/mnrasl/slv009 \\
\indent
Bianchi, E., Chandler, C. J., Ceccarelli, C., Codella, C., Sakai, N., L\'opez-Sepulcre, A., et al. (2020). FAUST I. The hot corino at the heart of the prototypical Class I protostar L1551 IRS5. MNRAS 498, L87–L92. doi:10.1093/mnrasl/slaa130 \\
\indent
Bisbas, T. G., Haworth, T. J., Barlow, M. J., Viti, S., Harries, T. J., Bell, T., et al. (2015). TORUS-3DPDR: a self-consistent code treating three-dimensional photoionization and photodissociation regions. MNRAS 454, 2828–2843. doi:10.1093/mnras/stv2156 \\
\indent
Caselli, P. and Ceccarelli, C. (2012). Our astrochemical heritage. ARA\&A 20, 56. doi:10.1007/ s00159-012-0056-x \\
\indent
Caselli, P., Vastel, C., Ceccarelli, C., van der Tak, F. F. S., Crapsi, A., and Bacmann, A. (2008). Survey of ortho-H2D+ (11,0-11,1) in dense cloud cores. A\&A 492, 703–718. doi:10.1051/0004-6361:20079009 \\
\indent
Ceccarelli, C., Bacmann, A., Boogert, A., Caux, E., Dominik, C., Lefloch, B., et al. (2010). Herschel
spectral surveys of star-forming regions. Overview of the 555-636 GHz range. A\&A 521, L22. doi:10. 1051/0004-6361/201015081 \\
\indent
Ceccarelli, C., Caselli, P., Bockel\'ee-Morvan, D., Mousis, O., Pizzarello, S., Robert, F., et al. (2014a). Deuterium Fractionation: The Ariadne's Thread from the Precollapse Phase to Meteorites and Comets Today. In Protostars and Planets VI, eds. H. Beuther, R. S. Klessen, C. P. Dullemond, and T. Henning. 859. doi:10.2458/azu\ uapress\ 9780816531240-ch037 \\
\indent
Ceccarelli, C., Caselli, P., Fontani, F., Neri, R., L\'opez-Sepulcre, A., Codella, C., et al. (2017). Seeds Of Life In Space (SOLIS): The Organic Composition Diversity at 300-1000 au Scale in Solar-type Star-forming Regions. ApJ 850, 176. doi:10.3847/1538-4357/aa961d \\
\indent
Ceccarelli, C., Caselli, P., Herbst, E., Tielens, A. G. G. M., and Caux, E. (2007). Extreme Deuteration and Hot Corinos: The Earliest Chemical Signatures of Low-Mass Star Formation. Protostars and Planets V , 47–62 \\
\indent
Ceccarelli, C., Dominik, C., L\'opez-Sepulcre, A., Kama, M., Padovani, M., Caux, E., et al. (2014b). Herschel Finds Evidence for Stellar Wind Particles in a Protostellar Envelope: Is This What Happened to the Young Sun? ApJ 790, L1. doi:10.1088/2041-8205/790/1/L1 \\
\indent
Ceccarelli, C., Hollenbach, D. J., and Tielens, A. G. G. M. (1996). Far-Infrared Line Emission from Collapsing Protostellar Envelopes. ApJ 471, 400. doi:10.1086/177978 \\
\indent
Ceccarelli, C., Viti, S., Balucani, N., and Taquet, V. (2018). The evolution of grain mantles and silicate dust growth at high redshift. MNRAS 476, 1371–1383. doi:10.1093/mnras/sty313 \\
\indent
Codella, C., Ceccarelli, C., Caselli, P., Balucani, N., Barone, V., Fontani, F., et al. (2017). Seeds of Life in Space (SOLIS). II. Formamide in protostellar shocks: Evidence for gas-phase formation. A\&A 605, L3. doi:10.1051/0004-6361/201731249 \\
\indent
Codella, C., Ceccarelli, C., Lee, C.-F., Bianchi, E., Balucani, N., Podio, L., et al. (2019). The HH 212 Interstellar Laboratory: Astrochemistry as a Tool To Reveal Protostellar Disks on Solar System Scales around a Rising Sun. ACS Earth and Space Chemistry 3, 2110–2121. doi:10.1021/acsearthspacechem. 9b00136 \\
\indent
Cruz-S\'aenz de Miera, F., K\'osp\'al, \'A., \'Abr\'aham, P., Liu, H. B., and Takami, M. (2019). Resolved ALMA
Continuum Image of the Circumbinary Ring and Circumstellar Disks in the L1551 IRS 5 System. ApJ 882, L4. doi:10.3847/2041-8213/ab39ea \\
\indent
De Simone, M., Ceccarelli, C., Codella, C., Svoboda, B. E., Chandler, C., Bouvier, M., et al. (2020). Hot Corinos Chemical Diversity: Myth or Reality? ApJ 896, L3. doi:10.3847/2041-8213/ab8d41 \\
\indent
De Simone, M., Codella, C., Testi, L., Belloche, A., Maury, A. J., Anderl, S., et al. (2017). Glycolaldehyde in Perseus Young Solar Analogs. A\&A 599, A121. doi:10.1051/0004-6361/201630049 \\
\indent
Dulieu, F., Congiu, E., Noble, J., Baouche, S., Chaabouni, H., Moudens, A., et al. (2013). How micron-sized dust particles determine the chemistry of our Universe. Scientific Reports 3, 1338. doi:10.1038/srep01338 \\
\indent
Fedele, D., Tazzari, M., Booth, R., Testi, L., Clarke, C. J., Pascucci, I., et al. (2018). ALMA continuum observations of the protoplanetary disk AS 209. Evidence of multiple gaps opened by a single planet. A\&A 610, A24. doi:10.1051/0004-6361/201731978 \\
\indent
Frank, A., Ray, T. P., Cabrit, S., Hartigan, P., Arce, H. G., Bacciotti, F., et al. (2014). Jets and Outflows from Star to Cloud: Observations Confront Theory. Protostars and Planets VI , 451–474.doi:10.2458/ azu uapress 9780816531240-ch020 \\
\indent
Galametz, M., Maury, A. J., Valdivia, V., Testi, L., Belloche, A., and Andr\'e, P. (2019). Low dust emissivities and radial variations in the envelopes of Class 0 protostars: possible signature of early grain growth. Astronomy and Astrophysics 632, A5. doi:10.1051/0004-6361/201936342 \\
\indent
Galv\'an-Madrid, R., Liu, H. B., Izquierdo, A. F., Miotello, A., Zhao, B., Carrasco-Gonz\'alez, C., et al. (2018). On the Effects of Self-obscuration in the (Sub)Millimeter Spectral Indices and the Appearance of Protostellar Disks. The Astrophysical Journal 868, 39. doi:10.3847/1538-4357/aae779 \\
\indent
Garrod, R. T., Widicus Weaver, S. L., and Herbst, E. (2008). Complex Chemistry in Star-forming Regions: An Expanded Gas-Grain Warm-up Chemical Model. ApJ 682, 283–302. doi:10.1086/588035 \\
\indent
Guillet, V., Pineau Des For\^ets, G., and Jones, A. P. (2011). Shocks in dense clouds. III. Dust processing and feedback effects in C-type shocks. A\&A 527, A123. doi:10.1051/0004-6361/201015973 \\
\indent
Herbst, E. and van Dishoeck, E. F. (2009). Complex Organic Interstellar Molecules. ARA\&A 47, 427–480. doi:10.1146/annurev-astro-082708-101654 \\
\indent
Higuchi, A. E., Sakai, N., Watanabe, Y., L\'opez-Sepulcre, A., Yoshida, K., Oya, Y., et al. (2018). Chemical Survey toward Young Stellar Objects in the Perseus Molecular Cloud Complex. ApJS 236, 52. doi:10.3847/1538-4365/aabfe9 \\
\indent
Ilee, J. D., Forgan, D. H., Evans, M. G., Hall, C., Booth, R., Clarke, C. J., et al. (2017). The chemistry of protoplanetary fragments formed via gravitational instabilities. MNRAS 472, 189–204. doi:10.1093/ mnras/stx1966 \\
\indent
Jørgensen, J. K., Belloche, A., and Garrod, R. T. (2020). Astrochemistry During the Formation of Stars. ARA\&A 58, 727–778. doi:10.1146/annurev-astro-032620-021927 \\
\indent
Lee, C.-F., Codella, C., Li, Z.-Y., and Liu, S.-Y. (2019). First Abundance Measurement of Organic Molecules in the Atmosphere of HH 212 Protostellar Disk. ApJ 876, 63. doi:10.3847/1538-4357/ab15db \\
\indent
Lee, C.-F., Ho, P. T. P., Li, Z.-Y., Hirano, N., Zhang, Q., and Shang, H. (2017a). A rotating protostellar jet launched from the innermost disk of HH 212. Nature Astronomy 1, 0152. doi:10.1038/s41550-017-0152 \\
\indent
Lee, C.-F., Li, Z.-Y., Ho, P. T. P., Hirano, N., Zhang, Q., and Shang, H. (2017b). Formation and Atmosphere of Complex Organic Molecules of the HH 212 Protostellar Disk. ApJ 843, 27. doi:10.3847/1538-4357/ aa7757 \\
\indent
Lefloch, B., Bachiller, R., Ceccarelli, C., Cernicharo, J., Codella, C., Fuente, A., et al. (2018). Astrochemical evolution along star formation: overview of the IRAM Large Program ASAI. MNRAS 477, 4792–4809. doi:10.1093/mnras/sty937 \\
\indent
Lesaffre, P., Pineau des For\^ets, G., Godard, B., Guillard, P., Boulanger, F., and Falgarone, E. (2013). Low-velocity shocks: signatures of turbulent dissipation in diffuse irradiated gas. A\&A 550, A106. doi:10.1051/0004-6361/201219928 \\
\indent
Miotello, A., Testi, L., Lodato, G., Ricci, L., Rosotti, G., Brooks, K., et al. (2014). Grain growth in the envelopes and disks of Class I protostars. Astronomy and Astrophysics 567, A32. doi:10.1051/ 0004-6361/201322945 \\
\indent
Okoda, Y., Oya, Y., Francis, L., Johnstone, D., Inutsuka, S.-i., Ceccarelli, C., et al. (2021). FAUST. II. Discovery of a Secondary Outflow in IRAS 15398-3359: Variability in Outflow Direction during the Earliest Stage of Star Formation? ApJ 910, 11. doi:10.3847/1538-4357/abddb1 \\
\indent
Oya, Y., Sakai, N., L\'opez-Sepulcre, A., Watanabe, Y., Ceccarelli, C., Lefloch, B., et al. (2016). Infalling- Rotating Motion and Associated Chemical Change in the Envelope of IRAS 16293-2422 Source A Studied with ALMA. ApJ 824, 88. doi:10.3847/0004-637X/824/2/88 \\
\indent
Podio, L., Garufi, A., Codella, C., Fedele, D., Bianchi, E., Bacciotti, F., et al. (2020). ALMA chemical survey of disk-outflow sources in Taurus (ALMA-DOT). II. Vertical stratification of CO, CS, CN, H2CO, and CH3OH in a Class I disk. A\&A 642, L7. doi:10.1051/0004-6361/202038952 \\
\indent
Sakai, N., Oya, Y., Higuchi, A. E., Aikawa, Y., Hanawa, T., Ceccarelli, C., et al. (2017). Vertical structure of the transition zone from infalling rotating envelope to disc in the Class 0 protostar, IRAS 04368+2557. MNRAS 467, L76–L80. doi:10.1093/mnrasl/slx002 \\
\indent
Sakai, N., Oya, Y., Sakai, T., Watanabe, Y., Hirota, T., Ceccarelli, C., et al. (2014a). A Chemical View of
Protostellar-disk Formation in L1527. ApJ 791, L38. doi:10.1088/2041-8205/791/2/L38 \\
\indent
Sakai, N., Sakai, T., Hirota, T., Watanabe, Y., Ceccarelli, C., Kahane, C., et al. (2014b). Change in the chemical composition of infalling gas forming a disk around a protostar. Nature 507, 78–80. doi:10.1038/nature13000 \\
\indent
Sakai, N. and Yamamoto, S. (2013). Warm Carbon-Chain Chemistry. Chemical Reviews 113, 8981–9015. doi:10.1021/cr4001308
Sheehan, P. D. and Eisner, J. A. (2017). Disk Masses for Embedded Class I Protostars in the Taurus Molecular Cloud. ApJ 851, 45. doi:10.3847/1538-4357/aa9990 \\
\indent
Skouteris, D., Vazart, F., Ceccarelli, C., Balucani, N., Puzzarini, C., and Barone, V. (2017). New quantum chemical computations of formamide deuteration support gas-phase formation of this prebiotic molecule. MNRAS 468, L1–L5. doi:10.1093/mnrasl/slx012 \\
\indent
Sta\"uber, P., Doty, S. D., van Dishoeck, E. F., and Benz, A. O. (2005). X-ray chemistry in the envelopes around young stellar objects. A\&A 440, 949–966. doi:10.1051/0004-6361:20052889 \\
\indent
Tabone, B., Cabrit, S., Bianchi, E., Ferreira, J., Pineau des For\^ets, G., Codella, C., et al. (2017). ALMA discovery of a rotating SO/SO2 flow in HH212. A possible MHD disk wind? A\&A 607, L6. doi:10. 1051/0004-6361/201731691 \\
\indent
Testi, L., Birnstiel, T., Ricci, L., Andrews, S., Blum, J., Carpenter, J., et al. (2014). Dust Evolution in Protoplanetary Disks. In Protostars and Planets VI, eds. H. Beuther, R. S. Klessen, C. P. Dullemond, and T. Henning. 339. doi:10.2458/azu\ uapress\ 9780816531240-ch015 \\
\indent
van’t Hoff, M. L. R., Harsono, D., Tobin, J. J., Bosman, A. D., van Dishoeck, E. F., Jørgensen, J. K., et al. (2020). Temperature Structures of Embedded Disks: Young Disks in Taurus Are Warm. ApJ 901, 166. doi:10.3847/1538-4357/abb1a2 \\
\indent
Walsh, C., Nomura, H., and van Dishoeck, E. (2015). The molecular composition of the planet-forming regions of protoplanetary disks across the luminosity regime. A\&A 582, A88. doi:10.1051/0004-6361/ 201526751 \\
\indent
Watanabe, N., Kimura, Y., Kouchi, A., Chigai, T., Hama, T., and Pirronello, V. (2010). Direct Measurements of Hydrogen Atom Diffusion and the Spin Temperature of Nascent H2 Molecule on Amorphous Solid Water. ApJ 714, L233–L237. doi:10.1088/2041-8205/714/2/L233 \\
\indent
Wirstr\"om, E. S. and Charnley, S. B. (2018). Revised models of interstellar nitrogen isotopic fractionation. MNRAS 474, 3720–3726. doi:10.1093/mnras/stx3030 \\
\indent
Yang, Y.-L., Sakai, N., Zhang, Y., Murillo, N. M., Zhang, Z. E., Higuchi, A. E., et al. (2021). The Perseus ALMA Chemistry Survey (PEACHES). I. The Complex Organic Molecules in Perseus Embedded Protostars. ApJ 910, 20. doi:10.3847/1538-4357/abdfd6 \\
\indent
Zhao, B., Caselli, P., Li, Z.-Y., Krasnopolsky, R., Shang, H., and Nakamura, F. (2016). Protostellar disc formation enabled by removal of small dust grains. MNRAS 460, 2050–2076. doi:10.1093/mnras/ stw1124 \\

%\bibliographystyle{alpha}
%\bibliography{sample}

\end{document}